\documentclass[twocolumn, prb, aps, 10pt, superscriptaddress]{revtex4-2}

\usepackage{graphicx}
\usepackage{amsmath}
\usepackage{amsfonts}
\usepackage{amssymb}
\usepackage{mathrsfs}
\usepackage{hyperref}
\hypersetup{colorlinks = true, allcolors = blue}
\usepackage[T1]{fontenc}
\usepackage[utf8]{inputenc}
\usepackage{textcomp}

\DeclareMathOperator{\arsinh}{arsinh}

\newcommand{\nn}{\nonumber\\}

\newcommand{\PP}[1]{\Phi_{#1}}
\newcommand{\PG}[1]{\Phi_{#1}^{\mathrm{G}}}
\newcommand{\PX}[1]{\Phi_{#1}^{\mathrm{X}}}
\newcommand{\Pp}[1]{\Phi_{#1}^{+}}
\newcommand{\Pm}[1]{\Phi_{#1}^{-}}
\newcommand{\Ppm}[1]{\Phi_{#1}^{\pm}}

\newcommand{\Cp}[1]{\Upsilon_{#1, +}}
\newcommand{\Cm}[1]{\Upsilon_{#1, -}}
\newcommand{\Cpm}[1]{\Upsilon_{#1, \pm}}

\newcommand{\Ce}[1]{\beta_{#1}}
\newcommand{\Cee}[1]{\beta^{(#1)}}

\newcommand{\Cepm}[1]{\beta_{#1, \pm}}

\newcommand{\wL}{\omega_{\mathrm{L}}}
\newcommand{\wX}{\omega_{\mathrm{X}}}
\newcommand{\wC}{\omega_{\mathrm{C}}}
\newcommand{\wXL}{\Delta \omega_{\mathrm{XL}}}
\newcommand{\wCL}{\Delta \omega_{\mathrm{CL}}}
\newcommand{\wCp}[1]{\omega_{#1, +}}
\newcommand{\wCm}[1]{\omega_{#1, -}}
\newcommand{\wCpm}[1]{\omega_{#1, \pm}}

\begin{document}

\title{Multiple wave packets running in the photon number-space}

\author{L.~Nimmesgern}
\affiliation{Lehrstuhl f\"{u}r Theoretische Physik III, Universit{\"a}t Bayreuth, 95440 Bayreuth, Germany}
\author{M.~Cygorek}
\affiliation{Heriot-Watt University, Edinburgh EH14 4AS, United Kingdom}
\author{A.~Mielnik-Pyszczorski}
\affiliation{Lehrstuhl f\"{u}r Theoretische Physik III, Universit{\"a}t Bayreuth, 95440 Bayreuth, Germany}
\author{D.~E.~Reiter}
\affiliation{Condensed Matter Theory, Department of Physics, TU Dortmund, 44221 Dortmund, Germany}
\author{A.~Vagov}
\affiliation{Lehrstuhl f\"{u}r Theoretische Physik III, Universit{\"a}t Bayreuth, 95440 Bayreuth, Germany}
\author{V.~M.~Axt}
\affiliation{Lehrstuhl f\"{u}r Theoretische Physik III, Universit{\"a}t Bayreuth, 95440 Bayreuth, Germany}

\begin{abstract}
If a two-level system coupled to a single-mode cavity is strongly driven by an external laser, instead of a continuous accumulation of photons in the cavity, oscillations in the mean photon number occur.
These oscillations correspond to peaks of finite width running up and down in the photon number distribution, reminiscent of wave packets in linear chain models.
A single wave packet is found if the cavity is resonant to the external laser.
Here, we show that for finite detuning multiple packet structures can exist simultaneously, oscillating at different frequencies and amplitudes.
We further study the influence of dissipative effects resulting in the formation of a stationary state, which depending on the parameters can be characterized by a bimodal photon number distribution.
While we give analytical limits for the maximally achievable photon number in the absence of any dissipation, surprisingly, dephasing processes can push the photon occupations towards higher photon numbers.
\end{abstract}

\maketitle

\section{Introduction}
\label{sec:introduction}
Ever since the dawn of quantum theory, there has been a strive towards controlling the quantum states of various matter systems such as atoms or molecules \cite{huang_controllability_1983, warren_coherent_1993, rabitz_whither_2000, chu_cold_2002, mabuchi_principles_2005, greilich_optical_2006, rabitz_focus_2009, ramsay_review_2010}.
Their interaction with electromagnetic fields is commonly used as a lever to achieve this goal.
More recently, control of the quantum state of light itself has attracted much attention, where now the matter systems take on the supportive role \cite{sayrin_real-time_2011, jenkins_controlled_2012, krastanov_universal_2015, ballantine_quantum_2021, ma_quantum_2021, cosacchi_deterministic_2022}.
With the methods of cavity quantum electrodynamics (cQED), nonclassical photonic states can be deliberately generated, including but not limited to Fock states \cite{michler_quantum_2000, varcoe_preparing_2000, hofheinz_generation_2008, zhou_field_2012, somaschi_near-optimal_2016, schweickert_-demand_2018, cosacchi_-demand_2020}, squeezed states \cite{eleuch_cavity_1999, lutterbach_production_2000, garces_strong_2016, joshi_qubit-flip-induced_2017, jabri_squeezed_2022} and Schr{\"o}dinger cat states \cite{brune_observing_1996, van_enk_entangled_2001, mirrahimi_dynamically_2014, ofek_extending_2016, cosacchi_schrodinger_2021}.
Besides the fundamental interest in nonclassical states as manifestations of the intricacies of quantum physics, they offer perspectives for technological advancements, e.g. through the use of in principle unlimited degrees of freedom to encode information \cite{ralph_quantum_2003, zoller_quantum_2005, lo_secure_2014, acin_quantum_2018, cosacchi_transiently_2020, schimpf_quantum_2021, vajner_quantum_2022, gao_quantum_2022, bozzio_enhancing_2022}.

In this work, we focus on the simplest model providing a platform to study cQED, namely a two-level system (TLS) coupled to a single-mode microcavity.
This model is capable of capturing the central features of a large variety of real systems, such as atoms, semiconductor quantum dots, or superconducting qubits.
Each realization is understood as sampling different regions of the parameter space \cite{cosacchi_n-photon_2022}.

We investigate novel highly nonclassical states that are characterized by multiple peaks in the photon number distribution, each of which constituting the center of a structure of finite width, which we will call a packet.
They exhibit rich dynamics with each peak oscillating at different frequencies and amplitudes.
We show that these states can be produced in simple fashion via strong continuous wave (cw) driving of the TLS.
As the result of dissipation, a stationary state will form after some time, which describes a similar bimodal distribution under the right circumstances.
The characteristics of both the dynamical behavior as well as the stationary state can be easily controlled by varying the driving strength and detuning the frequency of the external driving with respect to the resonance frequency of the cavity.
In addition, we discuss a remarkable mechanism that allows for occupation of photon numbers higher than in equivalent conservative systems if the dephasing of the TLS dominates the losses of the cavity.

\section{Theoretical model}
\label{sec:theory}

\begin{figure*}[t]
	\centering
	\includegraphics[scale = 1]{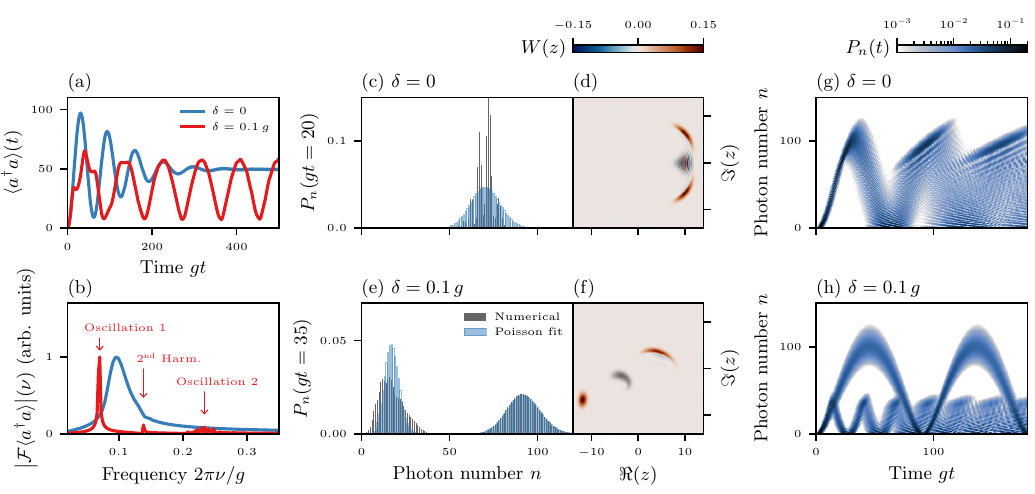}
	\caption{
	(a) Time-dependent mean photon number and (b) the absolute value of its Fourier transform calculated from solutions of Eq.~\eqref{eqn:schrodinger} using a driving strength of $f = 5\, g$ and detunings $\delta = 0$ (blue line) and $\delta = 0.1\, g$ (red line).
	(c) -- (f) Photon number distribution $P_{n}(t)$ (left panel) at fixed time and the corresponding Wigner function $W(z)$ obtained from solutions of Eq.~\eqref{eqn:schrodinger} with a driving strength of $f = 5\, g$ and detunings $\delta = 0$ (c), (d) and $\delta = 0.1\, g$ (e), (f).
		A Poisson distribution (blue boxes) is fit to each packet in the photon number distribution by calculating the mean and norm of the individual packets directly from the numerical solution (gray bars).
		(g), (h) Photon number distribution $P_{n}(t)$ resolved in both photon number $n$ as well as time $t$ obtained from solutions of Eq.~\eqref{eqn:schrodinger} with a driving strength of $f = 5\, g$ and detunings $\delta = 0$ (g) and $\delta = 0.1\, g$ (h).
	}
	\label{fig:numerical}
\end{figure*}

The TLS-cavity system is described by a Jaynes--Cummings model.
The Hamiltonian in the interaction picture with respect to the driving frequency $\wL$ reads
\begin{align}\label{eqn:hamiltonian}
	H
	={}& \hbar \wXL\, \sigma_{+} \sigma_{-}
	+ \hbar \wCL\, a^{\dagger} a\nn
	{}&+ \hbar g\, \bigl(a \sigma_{+} + a^{\dagger} \sigma_{-}\bigr)
	- \hbar f\, \bigl(\sigma_{+} + \sigma_{-}\bigr).
\end{align}
The ground state $\PG{}$ of the TLS is chosen as a state of energy zero, whereas the excited state $\PX{}$ has energy $\hbar \wX$.
$\wXL = \wX - \wL$ and $\wCL = \wC - \wL$ are the detunings between the external driving and the TLS and cavity respectively, where $\wC$ is the frequency of the cavity-mode.
The constant $g$ denotes the coupling strength between TLS and cavity, while $f$ is the external driving strength.
$a$ ($a^{\dagger}$) is the familiar photonic annihilation (creation) operator; $\sigma_{+}$ and $\sigma_{-}$ denote the ladder operators of the TLS satisfying fermionic anticommutation relations.
The basis of product states $\{\PG{n} \equiv \PG{} \otimes \Phi_{n},\, \PX{n} \equiv \PX{} \otimes \Phi_{n}\, |\, n \in \mathbb{N}_{0}\}$, named the bare state basis, spans the Hilbert space of the combined system, where $\Phi_{n}$ is the Fock state with photon number $n$.
For simplicity's sake we consider only cases of a resonantly driven TLS $\wXL = 0$.
Numerical studies indicate, that a finite value of $\wXL \lesssim \wCL$ does not noticeably change the results discussed in this work.
However, the novel phenomena that we shall present in this paper appear only for finite laser--cavity detuning.
We therefore have to allow for a finite $\wCL = \delta$.

As long as no dissipative effects are considered, we integrate the Schr{\"o}dinger equation
\begin{align}\label{eqn:schrodinger}
	i \hbar\, \dot{\Psi}(t) = H \Psi(t),
\end{align}
expanding the wave function $\Psi(t)$ in the bare state basis, by using a classic Runge-Kutta method (RK4).
Unless otherwise mentioned, the initial state is taken as the ground state without photons $\Psi(0) = \PG{0}$.

In Sec.~\ref{sec:dissipation} we present the behavior of the system under the influence of different dissipative effects.
All of them are assumed to be well-approximated as Markov processes, so that the dynamical evolution of the density matrix $\rho(t)$ can be calculated as the solution to the Liouville-von Neumann equation
\begin{align}\label{eqn:von_neumann}
	\dot{\rho}(t)
	={}& \frac{1}{i \hbar} \bigl[H, \rho(t)\bigr]\nn
	{}&+ \mathcal{L}_{\mathrm{cav}}\, \rho(t)
	+ \mathcal{L}_{\mathrm{RD}}\, \rho(t)
	+ \mathcal{L}_{\mathrm{PD}}\, \rho(t),
\end{align}
where the superoperators $\mathcal{L}_{\mathrm{cav}}$, $\mathcal{L}_{\mathrm{RD}}$ and $\mathcal{L}_{\mathrm{PD}}$ are of Lindblad form
\begin{subequations}\label{eqn:lindblad}
\begin{align}
	\label{eqn:lindblad_cav}
	\mathcal{L}_{\mathrm{cav}}\, \rho
	={}& \kappa\, \Bigl(a \rho a^{\dagger}
	- \frac{1}{2}\, a^{\dagger} a \rho
	- \frac{1}{2}\, \rho a^{\dagger} a\Bigr),
	\\
	\label{eqn:lindblad_rd}
	\mathcal{L}_{\mathrm{RD}}\, \rho
	={}& \gamma_{\mathrm{RD}}\, \Bigl(\sigma_{-} \rho \sigma_{+}
	- \frac{1}{2}\, \sigma_{+} \sigma_{-} \rho
	- \frac{1}{2}\, \rho \sigma_{+} \sigma_{-}\Bigr),
	\\
	\label{eqn:lindblad_pd}
	\mathcal{L}_{\mathrm{PD}}\, \rho
	={}& \frac{\gamma_{\mathrm{PD}}}{2}\, \Bigl(\sigma_{3} \rho \sigma_{3} - \rho\Bigr).
\end{align}
\end{subequations}
These terms model losses of the cavity, radiative decay and pure dephasing of the TLS, respectively.
We solve Eq.~\eqref{eqn:von_neumann} similar to Eq.~\eqref{eqn:schrodinger} by expanding the density matrix in the bare state basis and integrating the resulting system of differential equations by application of RK4.

\section{Numerical results}
\label{sec:numerics}
In order to obtain a clear picture of the dynamical behavior, we first investigate the solutions without any dissipation.
As will be shown in Sec.~\ref{sec:dissipation}, the same dynamical characteristics will be present transiently when dissipation is taken into account.

In Fig.~\ref{fig:numerical}(a) the time-resolved mean photon number $\langle a^{\dagger} a \rangle$ is shown in both a resonant ($\delta = 0$) and a non-resonant case ($\delta > 0$).
In the former, a harmonic damped oscillation can be seen.
Accordingly, its Fourier transform $\mathcal{F}\langle a^{\dagger} a \rangle$ [cf. Fig.~\ref{fig:numerical}(b)] consists of a single broad peak.
Due to the well-known \textit{collapse-and-revival} effect \cite{chough_nonlinear_1996}, the oscillation will reappear after some time (not shown in Figs.~\ref{fig:numerical}(a) and (b)).
We emphasize that the photon number will never exceed a finite maximal value even without dissipation.
This is remarkable since the system is continuously driven enabling an energy flow between the TLS and the laser.
Furthermore in the driven system all of the infinite steps of the Jaynes--Cummings ladder are coupled.

For finite but small detuning $\delta > 0$, a significantly different behavior is found.
The most prominent oscillation frequency is shifted with respect to that of $\delta = 0$ while the corresponding peak is of much lower spectral width.
Hence, the decay of the oscillation is only visible on a longer time scale.
In addition to small peaks at the second and third harmonic of this oscillation, contributions at a second higher fundamental frequency appear [cf. Fig.~\ref{fig:numerical}(b)].
Since it does not coincide with an integer multiple of the first frequency, this fact indicates the presence of two separate oscillating structures.

The photon number distribution
\begin{align}
	P_{n}(t)
	= \bigl|\bigl\langle \PG{n} \big| \Psi(t) \bigr\rangle\bigr|^{2}
	+ \bigl|\bigl\langle \PX{n} \big| \Psi(t) \bigr\rangle\bigr|^{2},
\end{align}
shown at fixed time in Fig.~\ref{fig:numerical}(c) for $\delta = 0$ and Fig.~\ref{fig:numerical}(e) for $\delta > 0$, reveals the two frequencies to correspond to two separate packets that oscillate independently from each other.
In contrast, only a single packet is present for $\delta = 0$ consistent with the single oscillation frequency of the mean photon number.
This is also in line with Ref.~\cite{chough_nonlinear_1996} where starting from the analytical solution, which is only available for $\delta = 0$ and without dissipation, a single oscillation of the mean photon number was approximately derived.

Distributions with multiple peaks obviously correspond to highly nonclassical states.
Yet, in some cases even the packets themselves can be classified as such, which can be shown by comparing them to best fits of Poisson distributions to the peak structures or identifying regions of negative values in the Wigner function \cite{barnett_methods_2002}
\begin{align}
	W(z) = \frac{2}{\pi} \mathrm{Tr}\Bigl((-1)^{a^{\dagger} a} D^{\dagger}(z) \rho_{\mathrm{phot}} D(z)\Bigr).
\end{align}
Here, $\rho_{\mathrm{phot}} = \mathrm{Tr}_{\{\PG{}, \PX{}\}}(\rho)$ denotes the photonic density matrix and $D(z) = \exp(z a^{\dagger} - z^{*} a)$ is the displacement operator.
Both the single packet in the resonant case as well as the packet of lower photon number in the detuned case clearly exhibit these characteristics of nonclassicality, whereas the packet of higher photon number is more akin to a coherent state [cf. Figs.~\ref{fig:numerical}(d) and \ref{fig:numerical}(f)].

Figs.~\ref{fig:numerical}(g) and \ref{fig:numerical}(h) visualize the time-dependence of $P_{n}(t)$ itself.
For $\delta = 0$, the initially present packet quickly disperses in agreement with the collapse of the oscillation of the mean photon number in Fig.~\ref{fig:numerical}(a).
Similar behavior is exhibited for $\delta > 0$ by the packet of lower oscillation amplitude, which incidentally corresponds to a higher oscillation frequency.
The packet of higher photon number shows a stable propagation for much longer times, providing further evidence that it behaves in a similar fashion to a single harmonic oscillator.
Indeed, this structure remains stable for more than 20 oscillation cycles for the particular set of parameters given in Fig.~\ref{fig:numerical}(h).

\begin{figure*}[t]
	\centering
	\includegraphics[scale=1]{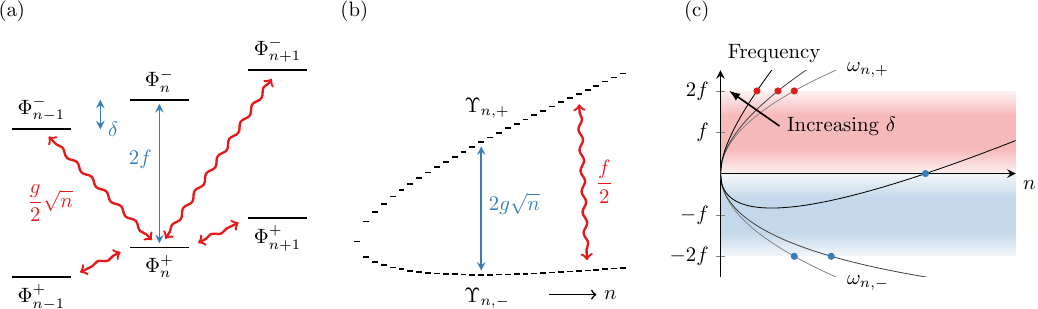}
	\caption{
		(a), (b) Energy level diagrams in the laser- and cavity-dressed state bases respectively in the case of a positive detuning $\delta > 0$.
		The blue arrows mark energy differences between states, whereas the red ones indicate possible transitions.
		(c) Sketch of the eigenfrequencies $\wCpm{n}$ of the cavity-dressed states as a function of the excitation number $n$ for multiple values of the detuning $\delta \geq 0$.
		The blue (red) background marks the area of wave-like propagation $0 \geq \wCm{n} \geq -2 f$ ($0 \leq \wCp{n} \leq 2 f$) of the wave packet corresponding to an initial state $\Pp{0}$ ($\Pm{0}$).
		The accordingly colored markers indicate the respective upper turning points.
	}
	\label{fig:analytical}
\end{figure*}

\section{Analytical results for strong driving}
\label{sec:strong_driving}
With the aim of gaining qualitative understanding of the dynamics, we analyze the situation of a strongly driven TLS.
Here, we concentrate on limiting cases where analytical results can be found.
We assume a cavity that is only slightly detuned, such that the hierarchy $\delta \ll g \ll f$ holds.

\subsection{Low excitation numbers}
\label{sec:strong_driving:low}
Since $f$ constitutes the dominant energy scale, it is natural to attempt a description in terms of the laser-dressed states (LDS)
\begin{align}\label{eqn:lds}
	\Ppm{} = \frac{1}{\sqrt{2}}\, \bigl(\PG{} \pm \PX{}\bigr),
\end{align}
which diagonalise the driving operator $-\hbar f \bigl(\sigma_{+} + \sigma_{-}\bigr)$.
The energies of the corresponding product states $\Ppm{n} = \Ppm{} \otimes \Phi_{n}$ are given by $\hbar \delta n \mp \hbar f$.
Via cavity coupling, transitions between states of neighboring photon numbers are induced, both at unchanged LDS as well as accompanying a transition $\Pp{} \leftrightarrow \Pm{}$, as sketched in Fig.~\ref{fig:analytical}(a).
The absolute values of the corresponding matrix elements are $\hbar g \sqrt{n} / 2$.

At the beginning of the dynamics and also after completion of a cycle in an oscillation of a packet only states with small photon numbers $n$ are occupied.
We concentrate on the regime where the following inequality holds for all occupied states with photon number $n$ in a given packet:
\begin{align}\label{eqn:lds_decoupling_n}
	n \ll \Bigr(\frac{4 f}{g}\Bigl)^{2}.
\end{align}
In this regime, the transitions $\Pp{n} \leftrightarrow \Pm{n \pm 1}$ become negligible, since the transition matrix elements are much smaller than the energy differences between these states.
However, this does not hold for the transitions $\Pp{n} \leftrightarrow \Pp{n \pm 1}$ and $\Pm{n} \leftrightarrow \Pm{n \pm 1}$.
Consequently, the two subspaces spanned by $\Pp{n}$ and $\Pm{n}$ can, to good approximation, be viewed as independent of each other.
In order to retain the effects of the cross-coupling $\Pp{n} \leftrightarrow \Pm{n \pm 1}$ perturbatively, effective Hamiltonian operators
\begin{align}\label{eqn:lds_decoupling_hamiltonian}
	H_{\mathrm{eff}}^{\pm}
	={}& \hbar \delta\, a^{\dagger} a \mp \hbar f
	\pm \frac{\hbar g}{2}\, \bigl(a + a^{\dagger}\bigr)
	\pm \frac{\hbar \vartheta}{2}\, \bigl(a - a^{\dagger}\bigr)^{2},
\end{align}
with
\begin{align}\label{eqn:lds_decoupling_theta}
	\vartheta = \Bigl(\frac{g}{2}\Bigr)^{2} \Bigl(\frac{1}{2 f + \delta} + \frac{1}{2 f - \delta}\Bigr),
\end{align}
can be derived, as detailed in Appendix~\ref{app:effective_hamiltonian}.
These describe the evolution within the respective subspaces.
The solutions to the corresponding Schr{\"o}dinger equations
\begin{align}\label{eqn:lds_decoupling_schrodinger}
	i \hbar\, \dot{\Psi}^{\pm}(t) = H_{\mathrm{eff}}^{\pm} \Psi^{\pm}(t)
\end{align}
with initial state $\Psi^{\pm}(0) = \Phi_{0}$ are characterized by mean photon numbers:
\begin{multline}\label{eqn:lds_decoupling_mean}
	\langle a^{\dagger} a \rangle^{\pm}(t)
	= \Bigl(\frac{g}{\delta}\Bigr)^{2}\, \frac{1 - \cos(\Omega^{\pm} t)}{2}\\
	+ \text{higher harmonics},
\end{multline}
where
\begin{align}
	\Omega^{\pm} = \delta \sqrt{1 \mp \frac{2 \vartheta}{\delta}}
\end{align}
(cf. Appendix~\ref{app:effective_hamiltonian}).
It is further noted, that the photon number distribution of $\Psi^{\pm}(t)$ will be approximately Poissonian since, for $\vartheta = 0$, which according to Eq.~\eqref{eqn:lds_decoupling_theta} is approached for strong driving, the solutions to Eq.~\eqref{eqn:lds_decoupling_schrodinger} are easily shown to be coherent states.
The superposition
\begin{align}
	\Psi(t) = \frac{1}{\sqrt{2}} \Bigl(\Pp{} \otimes \Psi^{+}(t) + \Pm{} \otimes \Psi^{-}(t)\Bigr)
\end{align}
is a consistent approximation to the solution of Eq.~\eqref{eqn:schrodinger} if and only if the maximal involved photon number fulfills Eq.~\eqref{eqn:lds_decoupling_n}, which is the case if
\begin{align}\label{eqn:lds_decoupling_det}
	|\delta| \gg \frac{g^{2}}{4 f}.
\end{align}
Thus, we find that, in this restricted parameter range, the photon number distribution consists of two packets of finite width oscillating independently, where the oscillations are equal in amplitude $g^{2} / \delta^{2}$ but differ in frequency $\Omega^{\pm}$.

\subsection{High excitation numbers}
\label{sec:strong_driving:high}
In contrast, $\hbar g \sqrt{n}$ constitutes the dominant energy scale at sufficiently high photon numbers $n$.
This regime can be reached when a packet climbs up high enough on the Jaynes--Cummings ladder.
In this case, a description in terms of the cavity-dressed states (CDS) is more appropriate, which are defined as the eigenstates of the undriven Jaynes--Cummings model:
\begin{align}
	\Bigl(\hbar \delta\, a^{\dagger} a
	+ \hbar g\, \bigl(a \sigma_{+} + a^{\dagger} \sigma_{-}\bigr)\Bigr) \Cpm{n}
	= \hbar \wCpm{n} \Cpm{n}.
\end{align}
For excitation numbers $n \gg 1$, these are given by
\begin{align}
	\Cpm{n} \approx \frac{1}{\sqrt{2}}\, \bigl(\PG{n} \pm \PX{n-1}\bigr)
\end{align}
with the corresponding eigenfrequencies
\begin{align}
	\wCpm{n} = \delta n \pm g \sqrt{n}.
\end{align}
The external driving induces transitions between states of neighboring excitation numbers $\Cp{n} \leftrightarrow \Cp{n \pm 1}$, $\Cm{n} \leftrightarrow \Cm{n \pm 1}$ and $\Cp{n} \leftrightarrow \Cm{n \pm 1}$ with approximately constant matrix elements of absolute value $\hbar f / 2$ for $n \gg 1$ [cf. Fig.~\ref{fig:analytical}(b)].

For high excitation numbers
\begin{align}\label{eqn:cds_decoupling_n}
	n \gg \Bigl(\frac{f}{4 g}\Bigr)^{2},
\end{align}
the energy differences between states $\Cp{n}$ and $\Cm{n \pm 1}$ are much larger than the respective coupling strengths.
Hence, the subspaces spanned by $\Cp{n}$ and $\Cm{n}$ are effectively decoupled, analogous the decoupling of the LDS in the preceding discussion.
In order to obtain understanding of the dynamical evolution within each subspace, we neglect any cross-coupling $\Cp{n} \leftrightarrow \Cm{n \pm 1}$ and determine the expansion coefficients $\Cepm{n}$ corresponding to an eigenstate at eigenfrequency $\lambda$ in the subspace spanned by $\Cpm{n}$ from
\begin{align}\label{eqn:cds_decoupling_mode}
	\hbar \wCpm{n}\, \Cepm{n}
	\mp \frac{\hbar f}{2}\, \bigl(\Cepm{n + 1} + \Cepm{n - 1}\bigr)
	&= \hbar \lambda\, \Cepm{n}.
\end{align}
Solutions of Eq.~\eqref{eqn:cds_decoupling_mode} approximate the structure of exact eigenstates at high excitation numbers.
According to a WKB approximation, which is presented in Appendix~\ref{app:wkb}, these show wave-like behavior in the range
\begin{align}\label{eqn:cds_decoupling_waverange}
	\lambda - f \leq \wCpm{n} \leq \lambda + f
\end{align}
and exponential decay outside thereof.
Thus, wave packets consisting of modes around some central eigenfrequency $\lambda_{0}$ will traverse this region, while being continually reflected at the turning points $N_{\pm}$, given by the boundaries of the range defined in Eq.~\eqref{eqn:cds_decoupling_waverange}.

With an initial state of $0$ photons, the wave function begins to evolve according to the laws of the LDS-decoupled regime.
Once the photon number is sufficiently high, the behavior dynamically transitions to that described by the CDS-decoupled regime.
In particular, those solutions of Eq.~\eqref{eqn:cds_decoupling_mode} that approximate eigenstates of $H$ with non-negligible overlap with the initial state will be relevant for the dynamical solution.
The spectrum of $\Ppm{0}$ is centered around
\begin{align}
	\langle H \rangle_{0}^{\pm} \equiv \langle \Ppm{0}\, |\, H \Ppm{0} \rangle = \mp \hbar f
\end{align}
with a width of
\begin{align}
	\sqrt{\langle H^{2} \rangle_{0}^{\pm} - \langle H \rangle_{0}^{\pm\, 2}}
	= \frac{\hbar g}{\sqrt{2}}.
\end{align}
In the case of strong driving $f \gg g$, we can therefore interpret $\Ppm{0}$ as a wave packet with $\lambda_{0} = \mp f$ and the initial state $\Psi(0) = \PG{0}$ as a superposition of both of these wave packets.
For $\delta = 0$, we find that $\Pp{0}$ ($\Pm{0}$) is constructed solely of modes in $\Cm{n}$ ($\Cp{n}$), since $\wCp{n} > 0$ ($\wCm{n} < 0$) for all $n$.
Due to the symmetry between $\wCp{n}$ and $\wCm{n}$ in the case of vanishing detuning, both wave packets exhibit equal characteristics.
In particular, they are reflected at the common lower and upper turning points
\begin{subequations}
\begin{align}
	N_{\pm}^{<}(\delta = 0) &= 0
	\\
	N_{\pm}^{>}(\delta = 0) &= \Bigl(\frac{2 f}{g}\Bigr)^{2},
\end{align}
\end{subequations}
in agreement with the results of Ref.~\cite{chough_nonlinear_1996}.
However, if $\delta$ is nonzero, this symmetry is broken leading to different behavior of the individual packets.
For sufficiently small values of $|\delta|$ the upper turning points are shifted slightly:
\begin{subequations}
\begin{align}
	N_{+}^{>}\Bigl(\delta > -\frac{g^{2}}{8 f}\Bigr)
	= \Bigl(\frac{g}{2 \delta}\Bigr)^{2}
	\Biggl(\sqrt{1 + \frac{8 f \delta}{g^{2}}} - 1\Biggr)^{2}
	\\
	\label{eqn:cds_decoupling_tpm_1}
	N_{-}^{>}\Bigl(\delta < \frac{g^{2}}{8 f}\Bigr)
	= \Bigl(\frac{g}{2 \delta}\Bigr)^{2}
	\Biggl(\sqrt{1 - \frac{8 f \delta}{g^{2}}} - 1\Biggr)^{2}.
\end{align}
\end{subequations}
In particular, for positive values of $\delta$, $N_{+}^{>}$ is increased, whereas $N_{-}^{>}$ is decreased and vice versa for $\delta < 0$.
If $\delta > g^{2} / 8 f$, the minimum of $\wCm{n}$ is greater than $-2 f$.
Similarly, if $\delta < -g^{2} / 8 f$, the maximum of $\wCp{n}$ is less than $2 f$.
As a result, $N_{-}^{>}$ and $N_{+}^{>}$ are given by the roots of $\wCm{n}$ or $\wCp{n}$ respectively:
\begin{align}\label{eqn:cds_decoupling_tpm_2}
	N_{-}^{>}\Bigl(\delta > \frac{g^{2}}{8 f}\Bigr)
	= N_{+}^{>}\Bigl(\delta < -\frac{g^{2}}{8 f}\Bigr)
	= \Bigl(\frac{g}{\delta}\Bigr)^{2},
\end{align}
coinciding with the amplitude of oscillation in the LDS-decoupled regime [cf. Eq.~\eqref{eqn:lds_decoupling_mean}].
This behavior of the turning points is sketched in Fig.~\ref{fig:analytical}(c).

The frequencies $\wCm{n}$ tend towards $+\infty$ in the limiting case $n \rightarrow \infty$ for any non-vanishing positive detuning.
As a consequence, there exist modes in $\Cm{n}$ at eigenvalues $\lambda \approx f$ that describe wave-like propagation between the turning points
\begin{subequations}
\begin{align}
	\label{eqn:cds_decoupling_tpm_tild_lower}
	\widetilde{N}_{-}^{<}
	={}& \Bigl(\frac{g}{\delta}\Bigr)^{2}
	\\
	\label{eqn:cds_decoupling_tpm_tild_upper}
	\widetilde{N}_{-}^{>}
	={}& \Bigl(\frac{g}{2 \delta}\Bigr)^{2}
	\Biggl(\sqrt{1 + \frac{8 f \delta}{g^{2}}} + 1\Biggr)^{2}.
\end{align}
\end{subequations}
The lower turning point lies at high excitation numbers $\widetilde{N}_{-}^{<} \gg 1$, which implies that the overlap of these eigenstates with $\Pm{0}$ is negligible.
Hence, these modes typically do not contribute to the dynamical solution.
However, if they overlap with the modes in $\Cp{n}$, the cross-coupling between CDS, which we neglected up to this point, leads to a coupling between the $\Cp{n}$- and $\Cm{n}$-modes.
It is noted, that a weak perturbation suffices for strong coupling between the modes, since they correspond to similar eigenvalues.
The overlap becomes non-negligible once $\widetilde{N}_{-}^{<} \approx N_{+}^{>}$, so that the $\Cm{n}$-modes are visible for parameters fulfilling
\begin{align}
	\delta \gtrsim \frac{g^{2}}{f}.
\end{align}
Whenever the packet arrives at $\widetilde{N}_{-}^{<} \approx N_{+}^{>}$, part of it will travel towards higher excitation numbers, as described by the $\Cm{n}$-modes, while the rest will propagate towards lower excitations numbers in accordance with the $\Cp{n}$-modes, resulting in a continuous split of the packet.

\begin{figure}[t]
	\centering
	\includegraphics[scale = 1]{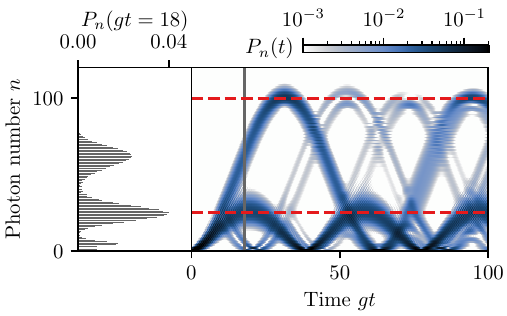}
	\caption{
		Photon number distribution $P_{n}(t)$ resolved in both photon number $n$ as well as time $t$ (right) and at fixed time $g t = 18$ (left) with a driving strength of $f = 5\, g$ and a detuning of $\delta = g^{2} / f = 0.2\, g$.
		The gray solid line indicates $g t = 18$, while the red dashed lines mark the intermediate turning point $\widetilde{N}_{-}^{<}$ given by Eq.~\eqref{eqn:cds_decoupling_tpm_tild_lower} and the upper turning point $\widetilde{N}_{-}^{>}$ given by Eq.~\eqref{eqn:cds_decoupling_tpm_tild_upper}.
	}
	\label{fig:split}
\end{figure}

With the roles of $\Cp{n}$ and $\Cm{n}$ reversed, the same phenomenon occurs if
\begin{align}
	\delta \lesssim -\frac{g^{2}}{f}.
\end{align}

\subsection{Comparison with numerical results}
\label{sec:comparison}
In typical situations, the photon number distribution consists of two oscillating packets, which are symmetric only if $\delta = 0$ and therefore cannot be distinguished in this case.
This qualitative picture imposed by the foregoing discussion is clearly supported by the numerical results presented in Sec.~\ref{sec:numerics}.
For $|\delta| \approx g^{2} / f$, more than two packets are expected owing to the continuous split at the intermediate turning point.
Similarly, this is confirmed by numerical calculations, as is shown in Fig.~\ref{fig:split}.

\begin{figure}[t]
	\centering
	\includegraphics[scale=1]{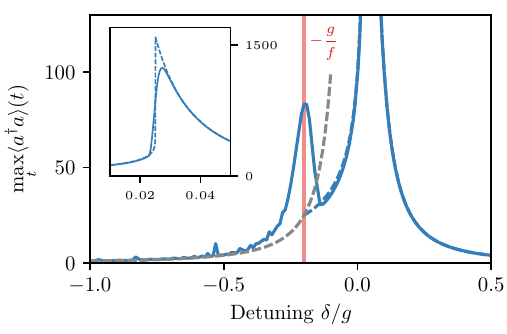}
	\caption{
		Maximum of the mean photon number (solid line) as a function of the detuning $\delta$ with a driving strength of $f = 5\, g$ compared to the turning points in the CDS-decoupled regime given by Eqs.~\eqref{eqn:cds_decoupling_tpm_1} and \eqref{eqn:cds_decoupling_tpm_2} (dashed line).
		The gray dashed line indicates the amplitude of the approximate mean photon number in the LDS-decoupled regime shown in Eq.~\eqref{eqn:lds_decoupling_mean}, whereas the red solid line shows the position $\delta = g^{2} / f$ of the area around which the split of packets occurs.
		All calculations are performed with initial state $\Psi(0) = \Pp{0}$.
		The inset figure shows the same results plotted for a lower range of $\delta$ and higher photon numbers.
	}
	\label{fig:max_mean_photonnumber}
\end{figure}

We now demonstrate that the analysis not only qualitatively, but also quantitatively reproduces the characteristics of the exact solution.
To this end, we perform calculations with initial state $\Psi(0) = \Pp{0}$, thereby isolating the packet in $\Pp{n}$ in the LDS-decoupled regime or the one in $\Cm{n}$ in the CDS-decoupled regime.
In Fig.~\ref{fig:max_mean_photonnumber}, we compare the maximum of the mean photon number with the turning points calculated in Sec.~\ref{sec:strong_driving}.
With the expected exception at $\delta \approx -g^{2} / f$, both are in excellent agreement.
As it turns out, the region around $\delta \approx -g^{2} / f$ appears to be the point of transition between the CDS- and LDS-decoupled solutions.

\section{Influence of dissipative effects}
\label{sec:dissipation}
In order to investigate, to what extend the previous results hold in realistic systems, which invariably experience some kind of dissipation, we turn our attention to solutions of the Liouville-von Neumann equation containing various dissipative contributions in Lindblad form, as detailed in Sec.~\ref{sec:theory}.

\begin{figure}[t]
	\centering
	\includegraphics[scale=1]{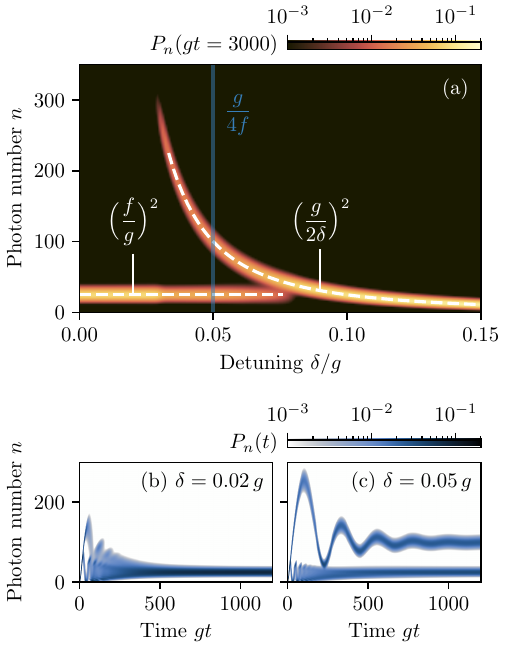}
	\caption{
		Photon number distribution $P_{n}(t)$ calculated by numerically integrating the Liouville-von Neumann equation~\eqref{eqn:von_neumann} with a driving strength of $f = 5\, g$ and cavity loss rate $\kappa = 0.01\, g$.
		Any dissipation of the TLS itself is neglected $\gamma_{\mathrm{RD}} = \gamma_{\mathrm{PD}} = 0$.
		Both the stationary distribution in dependence of the detuning $\delta$ (a) as well as its temporal evolution for values of the detuning of $\delta = 0.02\, g$ (b) and $\delta = 0.05\, g$ (c) are shown.
		In (a), the white dashed lines are graphs of the analytical expressions for the mean photon number of the packets given in Eqs.~\eqref{eqn:stat_n_detlow} and \eqref{eqn:stat_n_dethigh}, whereas the blue line indicates the parameter region $\delta = g^{2} / 4 f$ [cf. Eq.~\eqref{eqn:stat_det}] around which two packets are present in the stationary distribution.
	}
	\label{fig:cavity_losses}
\end{figure}

When isolating cavity losses, i.e. choosing $\gamma_{\mathrm{RD}} = \gamma_{\mathrm{PD}} = 0$, while $\kappa \neq 0$, the oscillations of the packets are damped, resulting in an eventual stationary state.
In general, the mean photon numbers of the individual packets in this stationary state differ from each other.
This provides the possibility of bimodal distributions, the existence of which were already reported in Ref.~\cite{cygorek_nonlinear_2017}.
However, only in the restricted set of parameters, given by
\begin{align}\label{eqn:stat_det}
	\delta \approx \frac{g^{2}}{4 f},
\end{align}
both packets are present in the stationary state.
For values of the detuning $\delta < g^{2} / 4 f$ only one packet with mean photon number
\begin{align}\label{eqn:stat_n_detlow}
	\langle a^{\dagger} a \rangle \approx \Bigl(\frac{f}{g}\Bigr)^{2}
\end{align}
and for $\delta > g^{2} / 4 f$ only one with mean photon number
\begin{align}\label{eqn:stat_n_dethigh}
	\langle a^{\dagger} a \rangle \approx \Bigl(\frac{g}{2 \delta}\Bigr)^{2}
\end{align}
are seen [cf. Fig.~\ref{fig:cavity_losses}(a)].
This is reflected in the transient behavior, as one of the packets experiences considerable decay in the respective parameter regions.
Illustrative examples of the dynamics in single- and dual-packet cases are shown in Figs.~\ref{fig:cavity_losses}(b) and \ref{fig:cavity_losses}(c).
It is to be noted, that the rules of thumb given in Eqs.~\eqref{eqn:stat_n_detlow} and \eqref{eqn:stat_n_dethigh} only hold for sufficiently low decay rates.
Obviously, for increasing values of $\kappa$, there will be a transition to a stationary state with $0$ photons in the limit $\kappa \rightarrow \infty$.

\begin{figure}[t]
	\centering
	\includegraphics[scale=1]{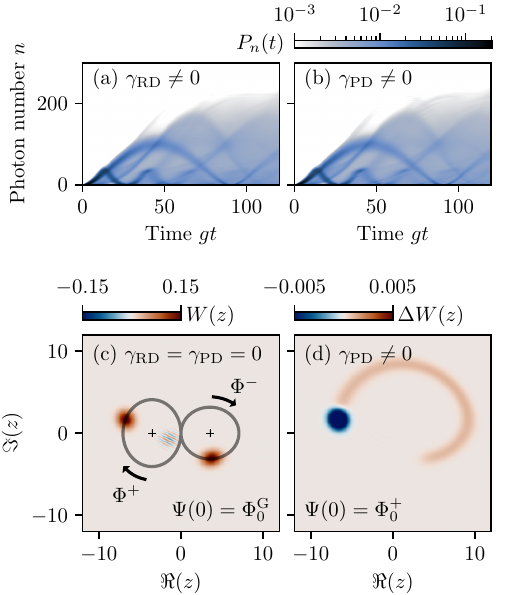}
	\caption{
		(a), (b) Dynamical evolution of the photon number distribution $P_{n}(t)$ with a driving strength of $f = 5\, g$ and a detuning $\delta = 0.1\, g$.
		Any dissipation is neglected with exception of radiative decay with $\gamma_{\mathrm{RD}} = 0.1\, g$ (a) or pure dephasing with $\gamma_{\mathrm{PD}} = 0.05\, g$ (b).
		(c) Wigner function $W(z)$ at fixed time $g t = 20$ starting from the initial state $\Psi(0) = \PG{0}$ in the LDS-decoupled regime ($f = 10\, g$, $\delta = 0.2\, g$) without any dissipation.
		(d) Difference between the Wigner functions with $\gamma_{\mathrm{PD}} = 0.005\, g$ and without any dissipation at fixed time $g t = 20$ starting from the initial state $\Psi(0) = \Pp{0}$ in the LDS-decoupled regime ($f = 10$, $\delta = 0.2\, g$).
	}
	\label{fig:qd_losses}
\end{figure}

Examining now the decay channels of the TLS while neglecting cavity losses, we find little qualitative difference between the effects of radiative decay and pure dephasing.
Examples of both cases are shown in Figs.~\ref{fig:qd_losses}(a) and \ref{fig:qd_losses}(b) respectively.
In contrast to the previously studied situation, the oscillations are not damped, but rather the packets themselves disperse until any discernible structure is lost.
Further, a number of additional packets are noticeable, which even reach higher photon numbers than those without dissipation.
Their origin can be understood in the LDS-decoupled case discussed in Sec.~\ref{sec:strong_driving:low}.
With the help of the phase space operators
\begin{subequations}
\begin{align}
	Q &= \frac{1}{\sqrt{2}} \Bigl(a + a^{\dagger}\Bigr)
	\\
	P &= -\frac{i}{\sqrt{2}} \Bigl(a - a^{\dagger}\Bigr)
\end{align}
\end{subequations}
the effective Hamiltonian operators that describe the photonic evolution within the isolated branches $\Ppm{}$ are written as [cf. Eq.~\eqref{eqn:lds_decoupling_hamiltonian}]
\begin{align}
	H_{\mathrm{eff}}^{\pm}
	&= \frac{\delta \mp 2 \vartheta}{2} P^{2}
	+ \frac{\delta}{2} \Bigl(Q \pm \frac{g}{\sqrt{2} \delta}\Bigr)^{2} + \mathrm{const}.
\end{align}
From this expression, we can immediately read off that parts of the Wigner function $W(z)$ corresponding to the TLS state $\Ppm{}$ will travel along ellipses around the center $z = \mp g / \sqrt{2} \delta$ with its width in $\Re(z)$ direction scaled by a factor of $\sqrt{1 \mp 2 \vartheta / \delta}$ with respect to its width in $\Im(z)$ direction.
This is visualized in Fig.~\ref{fig:qd_losses}(c) using a numerical calculation without any dissipation.
For the sake of clarity, we consider an initial state of $\Psi(0) = \Pp{0}$ for analyzing the influence of dephasing.
In the absence of any dissipation, the Wigner function resembles a Gaussian shape following the trajectory
\begin{align}
	z(t)
	= -\frac{g}{\sqrt{2} \delta} \Bigl(1 - \cos(\Omega^{+} t)\Bigr)
	- i \frac{g}{\sqrt{2} \Omega^{+}} \sin(\Omega^{+} t).
\end{align}
Pure dephasing induces transitions between the LDS $\Pp{} \leftrightarrow \Pm{}$ while the photonic state remains unchanged.
Hence, the original $\Pp{}$-solution is depleted leaving behind a trail of $\Pm{}$-states, where each part of this trail now orbits the center $z = g / \sqrt{2} \delta$ [cf. Fig.~\ref{fig:qd_losses}(d)].
By this mechanism, significantly higher photon numbers are reached.
For instance, sections which are generated at times $\Omega^{+} t = \pi$, i.e. when the original packet has reached its maximum $g^{2} / \delta^{2}$ [cf. Eq.~\eqref{eqn:lds_decoupling_mean}], will eventually reach $z = 4 g / \sqrt{2} \delta$ corresponding to a photon number of $|z|^{2} / 2 = 4 g^{2} / \delta^{2}$.
The appearance of additional packets is now attributed to situations in which large parts of the trail accumulate around similar photon numbers.
This interpretation is consistent with the fact, that the additional packets are sharply visible only during short periods of time.

\begin{figure}[t]
	\centering
	\includegraphics[scale=1]{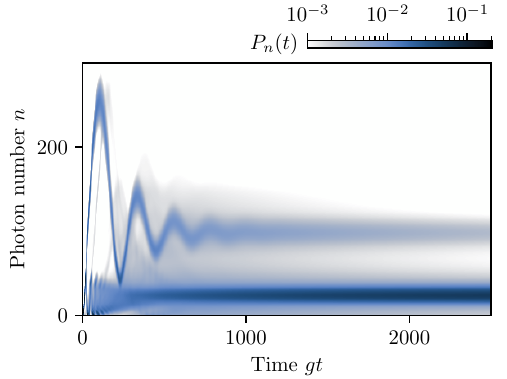}
	\caption{
		Photon number distribution $P_{n}(t)$ with a driving strength of $f = 5\, g$, a detuning $\delta = 0.05\, g$ and decay rates $\kappa = 0.01\, g$, $\gamma_{\mathrm{RD}} = 0.005\, g$ and $\gamma_{\mathrm{PD}} = 0.0025\, g$.
	}
	\label{fig:cavity_qd_losses}
\end{figure}

The effects of photonic and electronic dissipation combine in a straight-forward manner if both are taken into account concomitantly.
An example of this is shown in Fig.~\ref{fig:cavity_qd_losses}.
The packets induced by electronic dephasing as well as the main packets perform damped oscillations.
It is to be noted, that the upper packet is of considerably lower probability compared to the case of mere cavity losses.

\section{Conclusion}
\label{sec:conclusion}
We have studied the Jaynes--Cummings model with strong cw driving of the TLS.
We therein recovered the known result of a singular packet with oscillating mean photon number if the cavity is resonant to the external driving.
If the driving is slightly detuned with respect to the frequency of the cavity, two or more packets simultaneously oscillate with considerably different characteristics.

When taking into account various forms of dissipation, the oscillatory patterns remain present transiently.
Radiative decay and polarization dephasing of the TLS largely lead to the structure becoming indistinct whereas cavity losses damp the oscillations themselves.
We found that the latter results in the decay of one of two packets depending on the value of the detuning.
Only in a narrow transition region both packets remain stable under cavity losses, opening the possibility of bimodal stationary states.

We have presented a very simple way of preparing highly nonclassical photon states.
We can expect to find these states in numerous physical systems, owing to the universality of the model.
Our results might pave the way to photonic applications in particular new methods of quantum information processing that either exploit the bimodal stationary photon number distribution or the characteristic transient structures in the distribution to encode quantum information.
It should be noted that modern measurement techniques allow for a direct experimental monitoring of the photon number distribution \cite{schlottmann_exploring_2018, schmidt_photon-number-resolving_2018, helversen_quantum_2019} and therefore for a readout of information stored in the distribution.

\appendix
\section{Effective Hamiltonian in the strong driving limit}
\label{app:effective_hamiltonian}
The Hamiltonian in the LDS basis reads
\begin{align}
	H
	={}& \hbar \delta\, a^{\dagger} a
	+ \Bigl(\frac{\hbar g}{2}\, \bigl(a + a^{\dagger}\bigr) - \hbar f\Bigr)\, \mu_{3}\nn
	{}&+ \frac{\hbar g}{2}\, \bigl(a - a^{\dagger}\bigr) \bigl(\mu_{+} - \mu_{-}\bigr),
\end{align}
where $\mu_{+}$ and $\mu_{-}$ are the ladder operators defined by
\begin{gather}
\begin{aligned}
	\mu_{+} \Pp{} &= 0, &
	\mu_{+} \Pm{} &= \Pp{},
	\\
	\mu_{-} \Pp{} &= \Pm{}, &
	\mu_{-} \Pm{} &= 0,
\end{aligned}
\end{gather}
and $\mu_{3} = \mu_{+} \mu_{-} - \mu_{-} \mu_{+}$.
As described in Sec.~\ref{sec:strong_driving}, the cross-coupling, mediated by the operator $\mu_{+} - \mu_{-}$, only enacts a small effect on the dynamics at sufficiently low photon numbers.
We therefore seek a unitary transformation
\begin{align}
	U = \exp\bigl(i A\bigr),
\end{align}
with $A = A^{\dagger}$, such that
\begin{align}
	H_{\mathrm{eff}} = U H U^{\dagger}
\end{align}
is diagonal in the LDS-basis, thereby treating the influence of the cross-coupling perturbatively (for a detailed account of this method, see Ref.~\cite{cohen-tannoudji_atom-photon_2004}).
To second order in $g / f$ this is achieved by choosing
\begin{multline}
	A = i\, \Bigl(p_{1}\, a + q_{1}\, a^{\dagger}
	+ p_{2}\, a^{2} + q_{2}\, a^{\dagger\, 2}\\
	+ r_{2}\, \bigl(a^{\dagger} a + a a^{\dagger}\bigr)\Bigr)\, \mu_{+} + \mathrm{h.c.},
\end{multline}
where
\begin{subequations}
\begin{align}
	p_{1}
	={}& \frac{g}{2} \frac{1}{2 f + \delta}
	\\
	q_{1}
	={}& -\!\frac{g}{2} \frac{1}{2 f - \delta}
	\\
	p_{2}
	={}& \Bigl(\frac{g}{2}\Bigr)^{2} \frac{1}{(2 f + \delta) (f + \delta)}
	\\
	q_{2}
	={}& -\!\Bigl(\frac{g}{2}\Bigr)^{2} \frac{1}{(2 f - \delta) (f - \delta)}
	\\
	r_{2}
	={}& \Bigl(\frac{g}{2}\Bigr)^{2} \frac{1}{2 f}
	\Bigl(\frac{1}{2 f + \delta} - \frac{1}{2 f - \delta}\Bigr).
\end{align}
\end{subequations}
Using this expression we obtain up to an additive constant
\begin{align}
	H_{\mathrm{eff}}
	= H_{\mathrm{eff}}^{+}\, \mu_{+} \mu_{-}
	+ H_{\mathrm{eff}}^{-}\, \mu_{-} \mu_{+}
	+ \mathscr{O}\Bigl(\frac{g^{3}}{f^{3}}\Bigr),
\end{align}
where $H_{\mathrm{eff}}^{\pm}$ is given by Eq.~\eqref{eqn:lds_decoupling_hamiltonian}.

Since $H_{\mathrm{eff}}^{\pm}$ is quadratic in the photonic annihilation and creation operators, it can be readily diagonalised
\begin{align}
	H_{\mathrm{eff}}^{\pm} = \hbar \Omega^{\pm}\, b^{\dagger} b + \mathrm{const.},
\end{align}
such that $b$, $b^{\dagger}$ fulfill bosonic commutation relations.
We use the ansatz
\begin{align}\label{eqn:b_ansatz}
	b = \cosh(\chi^{\pm})\, a + \sinh(\chi^{\pm})\, a^{\dagger} + \zeta^{\pm},
\end{align}
with $\chi^{\pm},\, \zeta^{\pm} \in \mathbb{R}$, according to which
\begin{multline}
	b^{\dagger} b
	= \cosh(2 \chi^{\pm})\, a^{\dagger} a
	+ \zeta^{\pm} \exp(\chi^{\pm})\, \bigl(a + a^{\dagger}\bigr)\\
	+ \frac{1}{2} \sinh(2 \chi^{\pm})\, \bigl(a^{2} + a^{\dagger\, 2}\bigr)
	+ \mathrm{const.}
\end{multline}
By comparing coefficients with
\begin{multline}
	H_{\mathrm{eff}}^{\pm}
	= \hbar (\delta \mp \vartheta)\, a^{\dagger} a
	\pm \frac{\hbar g}{2}\, \bigl(a + a^{\dagger}\bigr)\\
	\pm \frac{\hbar \vartheta}{2}\, \bigl(a^{2} + a^{\dagger\, 2}\bigr)
	+ \mathrm{const.}
\end{multline}
and solving the resulting equations, we obtain the unique solution
\begin{subequations}
\begin{align}
	\Omega^{\pm}
	={}& \delta \sqrt{1 \mp \frac{2 \vartheta}{\delta}}
	\\
	\chi^{\pm}
	={}& \frac{1}{2} \arsinh\Bigl(\pm \frac{\vartheta}{\Omega^{\pm}}\Bigr)
	\\
	\zeta^{\pm}
	={}& \pm \frac{g}{2 \Omega^{\pm}} \exp(-\chi^{\pm}).
\end{align}
\end{subequations}
After solving Eq.~\eqref{eqn:b_ansatz} for $a$ and $a^{\dagger}$, the mean photon number under evolution of $H_{\mathrm{eff}}^{\pm}$ with initial condition $\PP{0}$ can be determined
\begin{multline}
	\langle a^{\dagger} a \rangle^{\pm}(t)
	= \Bigl(\frac{g}{\delta}\Bigr)^{2}\, \frac{1 - \cos(\Omega^{\pm} t)}{2}\\
	+ \frac{\vartheta}{2 \delta^{2}} \Bigl(\vartheta \pm \frac{g^{2}}{2 \Omega^{\pm}}\Bigr)\,
	\frac{1 - \cos(2 \Omega^{\pm} t)}{2}.
\end{multline}

\section{WKB analysis of a linear chain model}
\label{app:wkb}
We consider the equation
\begin{align}\label{eqn:lc_model}
	\omega(\varepsilon n)\, \Ce{n}
	+ \frac{\xi}{2}\, \bigl(\Ce{n+1} + \Ce{n-1}\bigr)
	= \lambda\, \Ce{n},
\end{align}
which recovers Eq.~\eqref{eqn:cds_decoupling_mode} after identifying $\omega(\varepsilon n) \rightarrow \wCpm{n}$, $\xi \rightarrow \mp f$ and $\Ce{n} \rightarrow \Cepm{n}$.
In the limit $\varepsilon \rightarrow 0$, corresponding to slowly varying frequencies $\omega$, we seek an asymptotic solution of the form
\begin{align}\label{eqn:lc_ansatz}
	\Ce{n}
	= \exp\Bigl(\frac{S(\varepsilon n)}{\varepsilon}\Bigr)
	\bigl(\Cee{0}(\varepsilon n) + \varepsilon \Cee{1}(\varepsilon n) + \mathscr{O}(\varepsilon^{2})\bigr),
\end{align}
where $S$ and $\Cee{j}$ are smooth functions.
In evaluating the expansion at neighboring excitation numbers, $S$ and $\Cee{j}$ are expanded in Taylor series
\begin{align}\label{eqn:lc_ansatz_expansion}
	\Ce{n \pm 1}
	= \exp\Bigl(\frac{S(\varepsilon n)}{\varepsilon} \pm S'(\varepsilon n)\Bigr)
	\bigl(\Cee{0}(\varepsilon n) + \mathscr{O}(\varepsilon)\bigr).
\end{align}
Inserting Eqs.~\eqref{eqn:lc_ansatz} and \eqref{eqn:lc_ansatz_expansion} into Eq.~\eqref{eqn:lc_model} results to leading order in the eikonal equation
\begin{align}\label{eqn:lc_eikonal}
	\frac{1}{2} \Bigl(\exp(S') + \exp(-S')\Bigr) = \frac{\lambda - \omega}{\xi}.
\end{align}
Since the right hand side of Eq.~\eqref{eqn:lc_eikonal} is real, $S'$ has to be either real, purely imaginary, or consist of a nonvanishing real part and a constant imaginary part of $\pi$.
These conform to the cases $(\lambda - \omega) / \xi < -1$, $-|\xi| < \lambda - \omega < |\xi|$ and $(\lambda - \omega) / \xi > 1$ respectively.
We thus conclude that $\Ce{n}$ exhibit wave-like behavior in the range given by
\begin{align}
	\lambda - |\xi| \leq \omega \leq \lambda + |\xi|
\end{align}
and exponential decay outside thereof.

In this work, we will not be interested in stating an explicit expression of $S$ or determining $\Cee{j}$.
The latter would be achieved by considering the higher order equations.
For a detailed discussion of the present method, see Ref.~\cite{holmes_introduction_2013} and references therein.

\bibliography{bib}

\end{document}